\documentclass[aps, pra, reprint, floatfix, showpacs]{revtex4-1}
\usepackage{graphicx}
\usepackage{amsmath,amssymb}
\usepackage{color}
\begin{document}
\title{Sympathetic cooling of fluorine atoms with ultracold atomic hydrogen}
\author{Maykel L.~Gonz\'alez-Mart\'{\i}nez}
\altaffiliation{Present address:
             Laboratoire Aim\'e Cotton, CNRS,
             Universit\'e Paris-Sud XI,
             ENS Cachan, B\^{a}t.\ 505, Campus d'Orsay,
             91405 Orsay, France}
\email{maykel.gonzalez-martinez@u-psud.fr}
\affiliation{Joint Quantum Centre (JQC) Durham/Newcastle,
             Department of Chemistry,
             Durham University, South Road,
             Durham DH1~3LE, United Kingdom}
\author{Jeremy M.~Hutson}
\email{J.M.Hutson@durham.ac.uk}
\affiliation{Joint Quantum Centre (JQC) Durham/Newcastle,
             Department of Chemistry,
             Durham University, South Road,
             Durham DH1~3LE, United Kingdom}
\date{\today}
\begin{abstract}
We consider the prospect of using ultracold hydrogen atoms for sympathetic
cooling of fluorine atoms to microkelvin temperatures. We carry out
quantum-mechanical calculations on collisions between cold F and H atoms in
magnetically trappable states and show that the ratio of elastic to inelastic
cross sections remains high across a wide range of temperatures and magnetic
fields. For F atoms initially in the spin-stretched state ($^2$P$_{3/2}$,
$f=m_f=+2$), sympathetic cooling appears likely to succeed from starting
temperatures around 1~K or even higher. This occurs because inelastic
collisions are suppressed by p-wave and d-wave barriers that are 600~mK and
3.2~K high, respectively. In combination with recent results on H + NH and H +
OH collisions [M. L. Gonz\'alez-Mart\'{\i}nez and J. M. Hutson, arXiv:1305.6282
(2013)], this establishes ultracold H atoms as a very promising and versatile
coolant for atoms and molecules that cannot be laser-cooled.
\end{abstract}
\pacs{37.10.De, 34.50.Cx
}
\maketitle
\section{Introduction}
Reaching the \emph{cold} ($T < 1$~K) and \emph{ultracold} ($T < 1$~mK) regimes
has triggered the revival of atomic physics.  In less than two decades, it has
led to the creation of Bose-Einstein condensates (BEC) \cite{MHAnderson:95,
CCBradley:95, KBDavis:95} and Fermi degenerate gases \cite{BDeMarco:99,
AGTruscott:01}, atom lasers \cite{MRAndrews:97}, quantized vortices
\cite{MRMatthews:99}, solitons \cite{SBurger:99, JDenschlag:00}, and optical
lattices \cite{IBloch:05}.  It has had a significant impact in high-precision
measurement and in the study of collective phenomena such as superfluidity and
superconductivity \cite{ODulieu:09a, LDCarr:09b, RVKrems:book09}.  In addition,
it has opened up the possibility of full control of atomic interactions and
scattering properties using external fields \cite{CChin:10}.

The atoms that can currently be cooled to the ultracold regime form only a
small part of the periodic table. They include alkali metals, some alkaline
earths, and relatively exotic species such as Yb, Cr, Dy, Er and metastable He,
all of which are amenable to laser Doppler cooling. However, there are many
other atoms that cannot yet be cooled to such temperatures, including
chemically important elements such as carbon, nitrogen, oxygen and the
halogens. Halogen atoms are particularly interesting; they are reactive species
that are often used as prototypes in the study of chemical reaction dynamics,
and ultracold halogen atoms would offer fascinating possibilities for exploring
chemical reactions in the cold and ultracold regime \cite{PFWeck:06c,
DHerschbach:09}.

Laser cooling of halogen atoms is not currently possible, since the UV lasers
needed to excite their lowest one-photon transitions are not available.
However, Doherty {\em et al.}\ \cite{WGDoherty:11} have recently used the
PhotoStop approach \cite{ATrottier:11} to trap Br atoms below 1~K at number
densities up to 10$^8$~cm$^{-3}$.  In addition, halogen atoms may be amenable to
Zeeman deceleration \cite{ENarevicius:07a}. However, such techniques by
themselves are unlikely to reach the true ultracold regime below 1~mK and a
second-stage cooling method is needed.

We have recently shown \cite{MLGonzalez-Martinez:13a} that sympathetic cooling
using spin-polarized ultracold atomic hydrogen offers a promising way to cool
prototype molecules to the ultracold regime. For NH, our calculations suggested
that sympathetic cooling with H atoms may be successful from a starting
temperature of 1~K or even higher; this contrasts with sympathetic cooling with
heavier atoms such as Li \cite{AOGWallis:09b} or Mg
\cite{AOGWallis:11a,MLGonzalez-Martinez:11}, which is predicted to succeed only
if the molecules can be precooled to 10 or 20 mK.  The general problem of
sympathetic cooling for species in electric \cite{HLBethlem:00b} or magnetic
traps \cite{JDWeinstein:98a} is that static traps can confine species only when
they are in \emph{low-field-seeking} states, and these states are never the
lowest state in the applied field. Collisions that transfer atoms or molecules
to the lower states release kinetic energy and usually eject both collision
partners from the trap. The key quantity that determines the feasibility of
sympathetic cooling is the ratio $\gamma$ between the cross-section for elastic
collisions (which produce thermalization) and that for inelastic collisions
(which cause trap loss). A common rule of thumb is that, for cooling to be
successful, this ratio needs to be at least 100 \cite{RdeCarvalho:99}.

In this paper we explore the use of sympathetic cooling with ultracold hydrogen
atoms for fluorine atoms in their ground state, $^2$P$_{3/2}$. We have modified
the MOLSCAT package \cite{JMHutson:MOLSCAT14, MLGonzalez-Martinez:07a} to carry
out quantum collision calculations between H and F atoms in a magnetic field,
including hyperfine interactions for both H and F. The H+F system is simple
enough that we can fully include all the potential curves that can contribute,
including the deeply bound $^1\Sigma^+$ ground state of HF, and also take full
account of hyperfine structure.


\section{Theory}
\label{sec:theory}

\subsection{Collision Hamiltonian}
\label{sec2:Heff} In this section we describe the general theory for collisions
between an atom A in a $^{2s_1+1}S$ state and an atom B in a state with orbital
angular momentum $l_2$ and electron spin $s_2$ in the presence of an external
magnetic field. We follow the convention of using lower-case letters for
operators and quantum numbers for the individual atoms, and capital letters for
those of the collision complex as a whole. The vector $\boldsymbol{R}$ joins
the centers of mass of the atoms, while $\boldsymbol{u}_r$ represents a unit
vector conjugate to the orbital angular momentum of atom B. Where necessary,
subscripts 1 and 2 refer to atoms A and B, respectively.

We solve the time-independent Schr\"odinger equation for the scattering wave
function $\Psi$ at energy $E$, $\hat{\mathcal{H}}\Psi = E\Psi$. The collision
Hamiltonian may be written
\begin{equation}
 \hat{\mathcal{H}} = - \frac{\hbar^2}{2\mu} R^{-1} \frac{d^2}{dR^2} R
                     + \frac{\hbar^2\hat{L}^2}{2\mu R^2}
                     + \hat{\mathcal{H}}_{1}
                     + \hat{\mathcal{H}}_{2}
                     + \hat{\mathcal{H}}_{12},
 \label{eq:Heff}
\end{equation}
where $R$ is the internuclear distance, $\hat{L}$ is the space-fixed operator
for the end-over-end rotation and $\mu$ is the reduced mass for the collision.
$\hat{\mathcal{H}}_1$ and $\hat{\mathcal{H}}_2$ describe the separated atoms,
while $\hat{\mathcal{H}}_{12}$ describes the interaction between them, with
contributions that include both the electronic potential $\hat{V}$ and the
dipolar interaction $\hat{\mathcal{H}}_\mathrm{dip}$ between the magnetic
moments of the two atoms.

In the general case of non-zero electronic and nuclear spins, the Hamiltonian
for an isolated atom B in a state $^{2s_2+1}l_2$ can be decomposed into
spin-orbit, hyperfine and Zeeman contributions,
\begin{equation}
\hat{\mathcal{H}}_2 =
\hat{\mathcal{H}}_\mathrm{so} + \hat{\mathcal{H}}_{\mathrm{hf},2} +
\hat{\mathcal{H}}_{\mathrm{Z},2}.
\end{equation}
In Russell-Saunders coupling, the spin-orbit term may be written
$\hat{\mathcal{H}}_\mathrm{so} = a_\mathrm{so} \hat{l}_2 \cdot \hat{s}_2$,
where $a_\mathrm{so}$ is the spin-orbit constant, and $\hat{l}_2$ and
$\hat{s}_2$ are the electronic orbital angular momentum and spin operators.
Corrections to this are required to handle deviations from Russell-Saunders
coupling \cite{EUCondon:35}.

The leading terms in $\hat{\mathcal{H}}_{\mathrm{hf},2}$ are the magnetic
dipole and electric quadrupole contributions. The magnetic dipole term may be
written $a_j \hat{\imath}_2 \cdot \hat{\jmath}_2$, where $\hat{\jmath}_2 =
\hat{l}_2+\hat{s}_2$ and the $a_j$ are the associated coupling constants
\cite{RETrees:53, JSMHarvey:65}.  The quadrupolar term (which exists only if
$i_2 \ge 1$, so is not present for $^1$H or $^{19}$F) may be written
$\hat{\mathcal{H}}_{\rm Q} = 2b_j \hat\imath_2\cdot\hat\jmath_2
(2\hat\imath_2\cdot\hat\jmath_2+1)$ \cite{RETrees:53, JSMHarvey:65}.

Finally, the Zeeman term arises from the interaction of the atomic magnetic
moment with the external magnetic field $\boldsymbol{B}$. It may be written
$\hat{\mathcal{H}}_{\rm Z,2} = -\hat{\mu}_2 \cdot \boldsymbol{B}$, where
$\hat{\mu}_2 = - g'_L\mu_\mathrm{B}\hat{l}_2 - g_S\mu_\mathrm{B} \hat{s}_2 +
g_{i2}\mu_\mathrm{N}\hat{\imath}_2$ \footnote{The effect of nuclear shielding
factors is extremely small and has been neglected in the present work.}, $g'_L
\approx 1$, $g_S \approx 2$ and $g_{i2}$ are the `corrected' orbital
\cite{JMBrown:Book03:p351}, electron and nuclear $g$-factors, and
$\mu_\mathrm{B}$ and $\mu_\mathrm{N}$ are the Bohr and nuclear magnetons.

The Hamiltonian for an isolated atom A in a $^{2s_1+1}S$ state is obtained from
that for B by setting $l_1 = 0$, and contains hyperfine and Zeeman
contributions only, so that $\hat{\mathcal{H}}_\mathrm{1} =
\hat{\mathcal{H}}_{\mathrm{hf},1} + \hat{\mathcal{H}}_{\rm Z,1}$.
$\hat{\mathcal{H}}_{\mathrm{hf},1}$ reduces to the Fermi contact interaction
$\hat{\mathcal{H}}_{\mathrm{F},1} = b_{\mathrm{F},1}\hat{s}_1 \cdot
\hat{\imath}_1$, with $b_{\mathrm{F},1}$ an effective coupling constant, and
$\hat{s}_1$ and $\hat{\imath}_1$ the electronic and nuclear spin operators. The
Zeeman term is the same as that for atom B, except that $\hat{\mu}_1 =
-g_S\mu_\mathrm{B}\hat{s}_1 + g_{i1}\mu_\mathrm{N}\hat{\imath}_1$.

The general expansion for the electronic interaction potential between two
atoms with arbitrary angular momentum has been derived by Krems, Groenenboom
and Dalgarno \cite{RVKrems:04b}.  First, since the operator for the total
interaction potential $\hat{V}(\boldsymbol{R},\boldsymbol{u}_r)$ is diagonal in
the total electronic spin $S$ ($\hat{S} = \hat{s}_1 + \hat{s}_2$) and its
space-fixed projection $M_S$, it can be decomposed into contributions
$\hat{V}^S(\boldsymbol{R},\boldsymbol{u}_r)$,
\begin{equation}
 \hat{V}(\boldsymbol{R},\boldsymbol{u}_r) = \sum^{s_1+s_2}_{S=|s_1-s_2|}
  \sum^{S}_{M_S=-S} |S M_S \rangle \langle S M_S|\,
  \hat{V}^S(\boldsymbol{R},\boldsymbol{u}_r).
 \label{eq:V}
\end{equation}
For interaction of an S-state atom A with an $l_2$-state atom B, there are
$l_2+1$ molecular electronic states for each $S$ with projections $|\Lambda| =
0,\cdots, l_2$ onto the internuclear axis. In the approximation that $l_2$ is
conserved at all values of $R$, these potentials can be represented by an
expansion in Legendre polynomials,
\begin{equation}
 \hat{V}^S(\boldsymbol{R},\boldsymbol{u}_r)
           = \sum_k \hat{V}^S_k(R) P_k(\boldsymbol{u}_R \cdot \boldsymbol{u}_r),
 \label{eq:VS}
\end{equation}
where $k = 0, 2,\ldots, 2l_2$, $\boldsymbol{u}_R \equiv (\theta,\phi)$ is a
unit vector in the direction of $\boldsymbol{R}$,
\begin{equation}
 \hat{V}^S_k(R) = \frac{(2k+1)}{(2l_2+1)} \sum^{l_2}_{\Lambda=-l_2}
                \frac{\langle l_2 \Lambda k 0 | l_2 \Lambda\rangle}
                {\langle l_2 0 k 0 | l_2 0 \rangle}
                \hat{V}^S_\Lambda(R),
 \label{eq:VSk}
\end{equation}
and $\langle j_1 m_1 j_2 m_2 | j m \rangle$ is a Clebsch-Gordan coefficient.
For the specific case where atom B is a halogen, $l_2 = 1$, and
Eq.~\eqref{eq:VSk} reduces to one isotropic ($k = 0$) and one anisotropic ($k =
2$) component for each $S$,
\begin{eqnarray}
 \hat{V}^S_0(R) &=&
  \frac{1}{3}\left[\hat{V}^S_\Sigma(R) + 2\hat{V}^S_\Pi(R)\right],\\
 \hat{V}^S_2(R) &=&
  \frac{5}{3}\left[\hat{V}^S_\Sigma(R) - \hat{V}^S_\Pi(R)\right].
 \label{eq:VS02}
\end{eqnarray}

The direct dipolar interaction between the magnetic moments of the two atoms
may be written \cite{JMBrown:Book03}
\begin{equation}
 \hat{\mathcal{H}}_\mathrm{dip}=-\sqrt{6}(\mu_0/4\pi)R^{-3}
  \mathrm{T}^2(\hat{\mu}_1,\hat{\mu}_2) \cdot \mathrm{T}^2(\boldsymbol{u}_R),
 \label{eq:Hdip}
\end{equation}
where $\mu_0$ is the magnetic permeability of free space and $T^k$ represent a
spherical tensor of rank $k$.

In the present work, we assume that all the atomic coupling constants are
independent of the internuclear distance $R$.  Any variation of these constants
could be introduced as an additional term in the interaction operator
$\hat{\mathcal{H}}_{12}$, as has been done, for example, for the $R$-dependence
of hyperfine interactions for the interactions between alkali-metal atoms and
closed-shell atoms \cite{Zuchowski:RbSr:2010, Brue:AlkYb:2013}. Such effects
might include, for example, the distance-dependence of the spin-orbit coupling
constant $a_\mathrm{so}$, or a nuclear quadrupole interaction in atom A (if
$i_1 \ge 1$) as a result of the breakdown of its spherical symmetry when the
two atoms approach one another.

\subsection{Coupled-channel equations}
\label{sec2:coupled-eqs} We solve the quantum-mechanical scattering problem
using the coupled-channel method. The total wave function is first expanded in
a set of $N$ conveniently chosen basis functions $\left|a\right>$
\begin{equation}
 \Psi(R,\xi) = R^{-1} \sum_{a} \chi_a(R) \left|a\right>,
\end{equation}
where $\xi$ is a collective variable including all coordinates except $R$, and
$a$ is the set of quantum numbers that label the basis functions.  Each
different combination of quantum numbers $a$ defines a \emph{channel}. A set of
coupled differential equations for the \emph{channel functions} $\chi_a(R)$ is
then obtained by substituting $\Psi(R,\xi)$ into the time-independent
Schr\"odinger equation to yield,
\begin{equation}
 \frac{d^2 \chi_a}{dR^2}
              = \sum_{a'} \left(W_{aa'} - \epsilon\delta_{aa'}\right) \chi_{a'},
 \label{eq:coupled-eqs}
\end{equation}
where $\delta_{ij}$ is the Kronecker delta, $\epsilon = 2\mu E/\hbar^2$ is a
scaled energy and
\begin{equation}
 W_{aa'}(R) = \frac{2\mu}{\hbar^2}
              \big\langle a \big| \hat{\mathcal{H}}_1 + \hat{\mathcal{H}}_2
              + \hat{\mathcal{H}}_{12}
              + \frac{\hbar^2\hat{L}^2}{2\mu R^2} \big|a'\big\rangle.
 \label{eq:waa'}
\end{equation}

The coupled equations \eqref{eq:coupled-eqs} are solved by propagating a
complete set of independent solution vectors from $R_\mathrm{min}$, deep in the
inner classically forbidden region, to $R_\mathrm{max}$, large enough that the
effects of the interaction potential have died off.

\subsection{Basis sets and matrix elements}
\label{sec2:basis+aHa'} The basis functions may be written $|a\rangle \equiv
|\alpha_1\rangle |\alpha_2\rangle |L M_L\rangle$, where $|\alpha_1\rangle$ and
$|\alpha_2\rangle$ are basis functions for atoms A and B, and $M_X$ (or $m_X$)
denotes the projection of the angular momentum $X$ along the axis of the
magnetic field. We use two different basis sets, with common $|\alpha_1\rangle
= |i_1 m_{i1}\rangle |s_1 m_{s1}\rangle$. The difference occurs in the
description of atom B, for which we used both a totally uncoupled basis set
$|i_2 m_{i2}\rangle |l_2 m_{l2}\rangle |s_2 m_{s2}\rangle$ and an
`$ls$-coupled' basis set $|i_2 m_{i2}\rangle |(l_2 s_2)j_2 m_{j2}\rangle$. We
have verified that scattering calculations in these two basis sets give
identical results.

In a static magnetic field, the only conserved quantities are the projection
$M_\mathrm{tot}$ of the total angular momentum and the total parity $P$ of the
system, which are explicitly $M_\mathrm{tot} = m_{i1} + m_{s1} + m_{i2} +
m_{l2} + m_{s2} + M_L = m_{i1} + m_{s1} + m_{i2} + m_{j2} + M_L$ and $P = p_1
p_2 (-1)^L$, where $p_1$ and $p_2$ are the parities of atoms A and B. The
coupled equations are therefore constructed and solved separately for each
combination of $M_{\rm tot}$ and $P$.

Constructing the coupled equations \eqref{eq:coupled-eqs} requires evaluating
the matrix elements of all the terms in $\hat{\mathcal{H}}_{1}$,
$\hat{\mathcal{H}}_{2}$ and $\hat{\mathcal{H}}_{12}$ in the chosen basis set.
The resulting matrix elements are given in Appendix~\ref{app:matrix_elements}.
The parameters for $^{19}$F hyperfine Hamiltonian were taken from
Ref.~\cite{JSMHarvey:65}.

\subsection{Interaction potentials} \label{sec2:PECs}
\begin{figure}
 \includegraphics[width=85mm]{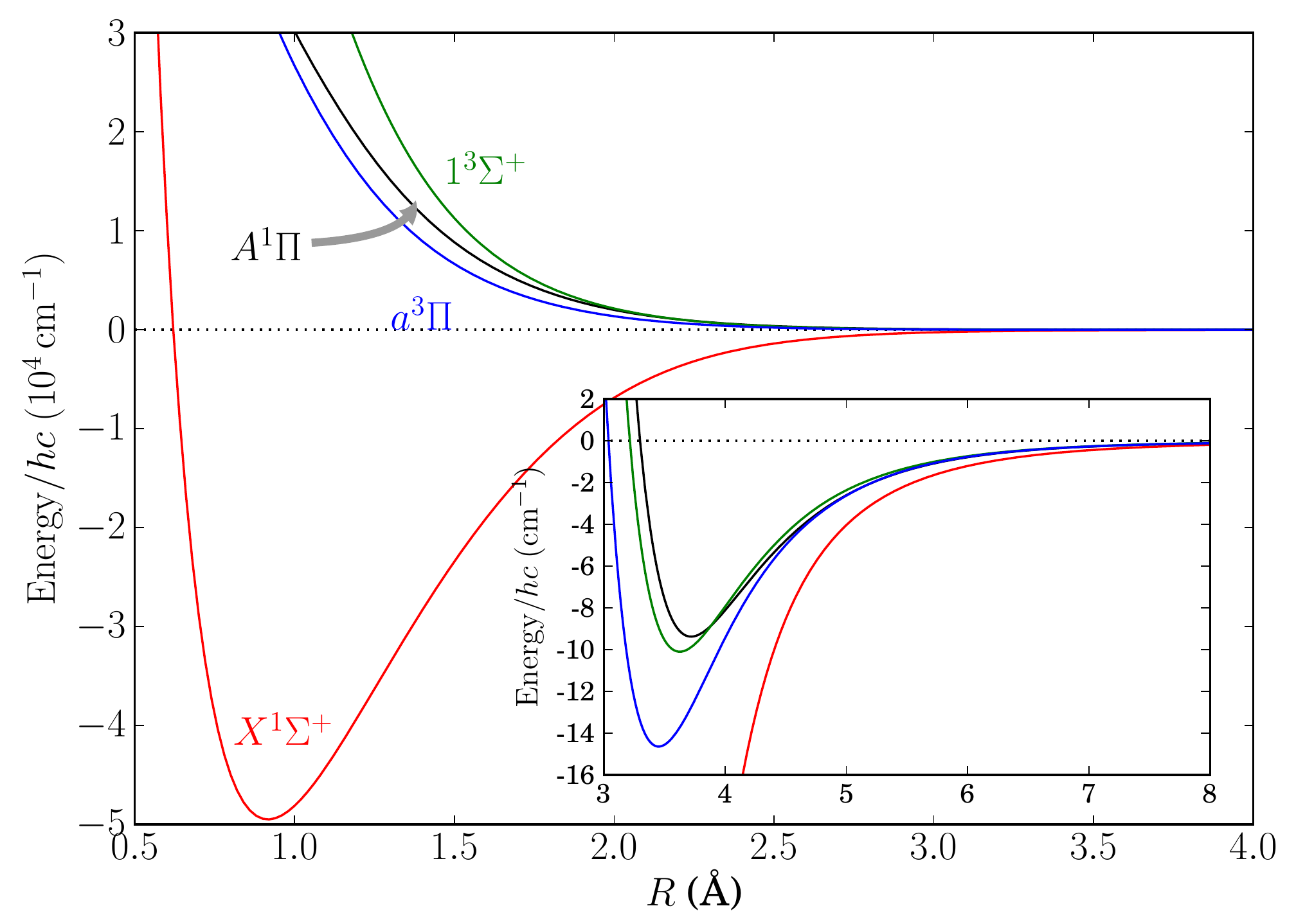}
 \caption{(Color online) Ground ($X{^1\Sigma^+}$) and excited ($A{^1\Pi}$,
   $1{^3\Sigma^+}$,
   $a{^3\Pi}$) electronic states of H+F.  The inset shows the relatively shallow
   Van der Waals wells for the excited states.
 \label{fig:PECs}}
\end{figure}

The interaction between the H($^2$S$_{1/2}$) and F($^2$P) atoms gives rise to
four electronic states, shown in Fig.~\ref{fig:PECs}. The $X{^1\Sigma^+}$ state
is strongly chemically bound, with a well about 50,000~cm$^{-1}$ (over 6~eV)
deep near $R=0.9$~\AA. By contrast, the $A{^1\Pi}$, $1{^3\Sigma^+}$ and
$a{^3\Pi}$ excited states are weakly bound, with long-range Van der Waals wells
between 8 and 15~cm$^{-1}$ deep at about 3.5~\AA.

The potential curve for the ground state of HF has been determined very
accurately by Coxon and Hajigeorgiou \cite{JACoxon:12} from a global
least-squares fit of all the spectroscopic data available in the literature.
However, this curve is not strictly the curve for the pure ${^1\Sigma^+}$ state:
close to dissociation, it should be interpreted as the lowest eigenvalue of a
fixed-$R$ electronic Hamiltonian, including spin-orbit coupling, which
correlates with the $^2$P$_{3/2}$ ground state of F. We have therefore used
curves for the three excited states obtained by Brown and Balint-Kurti
\cite{ABrown:00a} from multi-reference configuration interaction (MRCI)
calculations, and determined the ${^1\Sigma^+}$ potential so that the lowest
eigenvalue of the $\Omega=0$ matrix \cite{Aubert:1998}
\begin{equation}
 \left(\begin{array}{cc}
  V^\mathrm{diab}_1-\frac{1}{2}a_\mathrm{so} &
   \frac{1}{\sqrt{2}}a_\mathrm{so} \\
  \frac{1}{\sqrt{2}}a_\mathrm{so} & V^\mathrm{diab}_2
  \end{array}\right),
\end{equation}
matches the ground-state curve of Coxon and Hajigeorgiou \cite{JACoxon:12}.
Here, $1 \equiv {^3\Pi_{0,e}}$ and $2 \equiv {^1\Sigma_{0,e}}$ as seen in
Fig.~\ref{fig:PECs}.  In this calculation, we neglected the $R$-dependence of
the spin-orbit Hamiltonian. The resulting points for all 4 curves were
inter-/extrapolated using the reproducing kernel Hilbert space (RKHS) method
\cite{T-SHo:96a,PSoldan:00,T-SHo:00}, with the $C_6$, $C_8$ and $C_{10}$
dispersion coefficients constrained to match the scaled values of
Ref.~\cite{OZatsarinny:09}.

\subsection{Numerical methods}
\label{sec2:xsections}

The coupled equations \eqref{eq:coupled-eqs} are constructed in the primitive
basis sets described in section \ref{sec2:basis+aHa'}. The basis set includes
partial waves up to $L_\mathrm{max} = 8$, which is well converged. There are
typically 100 to 200 channels, depending on $M_{\rm tot}$ and parity.

The coupled equations are solved using the hybrid log-derivative Airy method of
Alexander and Manolopoulos \cite{MHAlexander:87}, using a fixed-step-size
log-derivative propagator for $0.5 \le R \le 25$~\AA, with $\Delta R =
0.005$~\AA, and a variable-step-size Airy propagator for $25 \le R \le
10,000$~\AA. The wavefunction log-derivative matrix is transformed at
$R_\mathrm{max}=10,000$~\AA\ into a basis set in which $\mathcal{H}_1$,
$\mathcal{H}_2$ and $\hat L^2$ are diagonal \cite{MLGonzalez-Martinez:07a}, and
the transformed channel functions are matched to the standard scattering
boundary conditions \cite{BRJohnson:73}. This gives the S-matrix from which the
state-to-state cross sections and scattering lengths are calculated.

\section{Results and discussion}
\label{sec:results+discussion}

\subsection{Atomic hyperfine and Zeeman levels}
\label{sec2:elevels}

\begin{figure}
 \includegraphics[width=85mm]{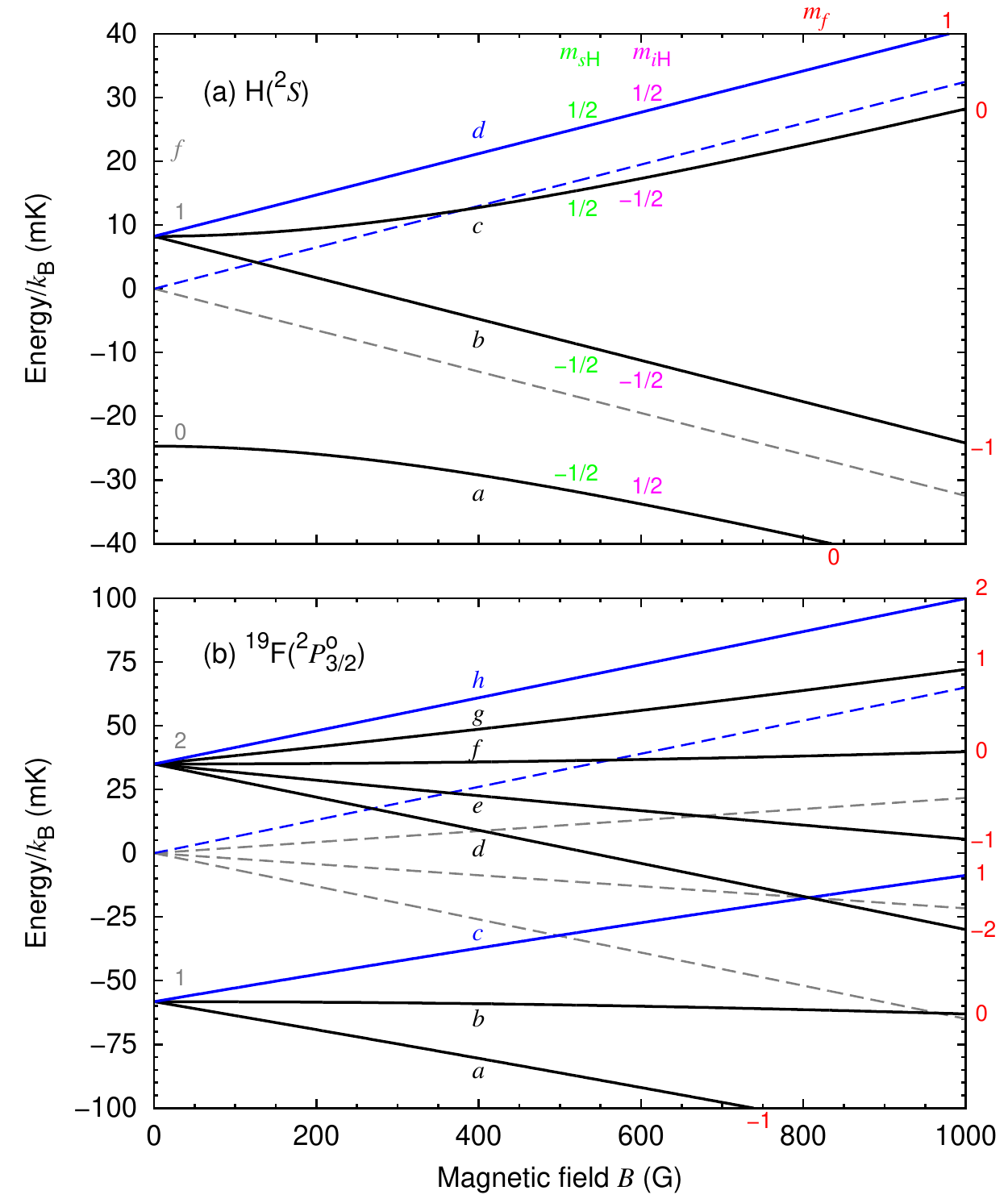}
\caption{(Color online) Magnetic-field dependence of the energy levels for: (a)
H($^2$S) and (b) $^{19}$F($^2P^\mathrm{o}_{3/2}$).  Solid (dashed) lines
correspond to the inclusion (exclusion) of hyperfine terms.  Scattering
calculations were carried out for the magnetically trappable states highlighted
in blue.
 \label{fig:elevels}}
\end{figure}

Figure \ref{fig:elevels}(a) shows the energy levels of the H atom in a magnetic
field. For sympathetic cooling, we consider collisions with atoms that are
magnetically trapped in the doubly-polarized state $\left|
^2\mathrm{S}_{1/2},\; m_{s\mathrm{H}} = +\frac{1}{2},\; m_{i\mathrm{H}} =
+\frac{1}{2} \right>$, designated H$_d$ and shown as a solid blue line in
Fig.~\ref{fig:elevels}(a).

Fig.~\ref{fig:elevels}(b) shows the corresponding energy levels of the $^{19}$F
atom. We have carried out calculations on H+F collisions for the two
low-field-seeking initial states shown as solid blue lines. These correspond to
$\left| ^2\mathrm{P}_{3/2},\; f_\mathrm{F} = 1,\; m_{f\mathrm{F}} = +1 \right>$
and $\left| ^2\mathrm{P}_{3/2},\; f_\mathrm{F} = 2,\; m_{f\mathrm{F}} = +2
\right>$, and are designated F$_c$ and F$_h$ respectively.

Hydrogen atoms cannot yet be laser-cooled, because the laser required to drive
the 2$^2$P $\leftarrow$ 1$^2$S transition (Lyman-$\alpha$) is not available.
Nevertheless, magnetically trapped hydrogen atoms in the state H$_d$ have been
produced at temperatures of 40 to 100~mK and densities up to $3\times10^{14}$
cm$^{-3}$ by purely cryogenic methods \cite{HFHess:87, RvanRoijen:88}, and then
evaporatively cooled to produce a Bose-Einstein condensate (BEC) of $10^9$ atoms
at a temperature around 50~$\mu$K and densities between $10^{14}$ and
$5\times10^{15}$ cm$^{-3}$ \cite{DGFried:98}. In addition, Zeeman deceleration
and magnetic trapping of hydrogen has recently been demonstrated
\cite{NVanhaecke:07,SDHogan:07,SDHogan:08,SDHogan:09}, although at higher
temperatures and lower number densities. For sympathetic cooling purposes BEC is
unnecessary, but the large densities and cloud sizes achievable by cryogenic
methods are very valuable.

\begin{figure}
 \includegraphics[width=85mm]{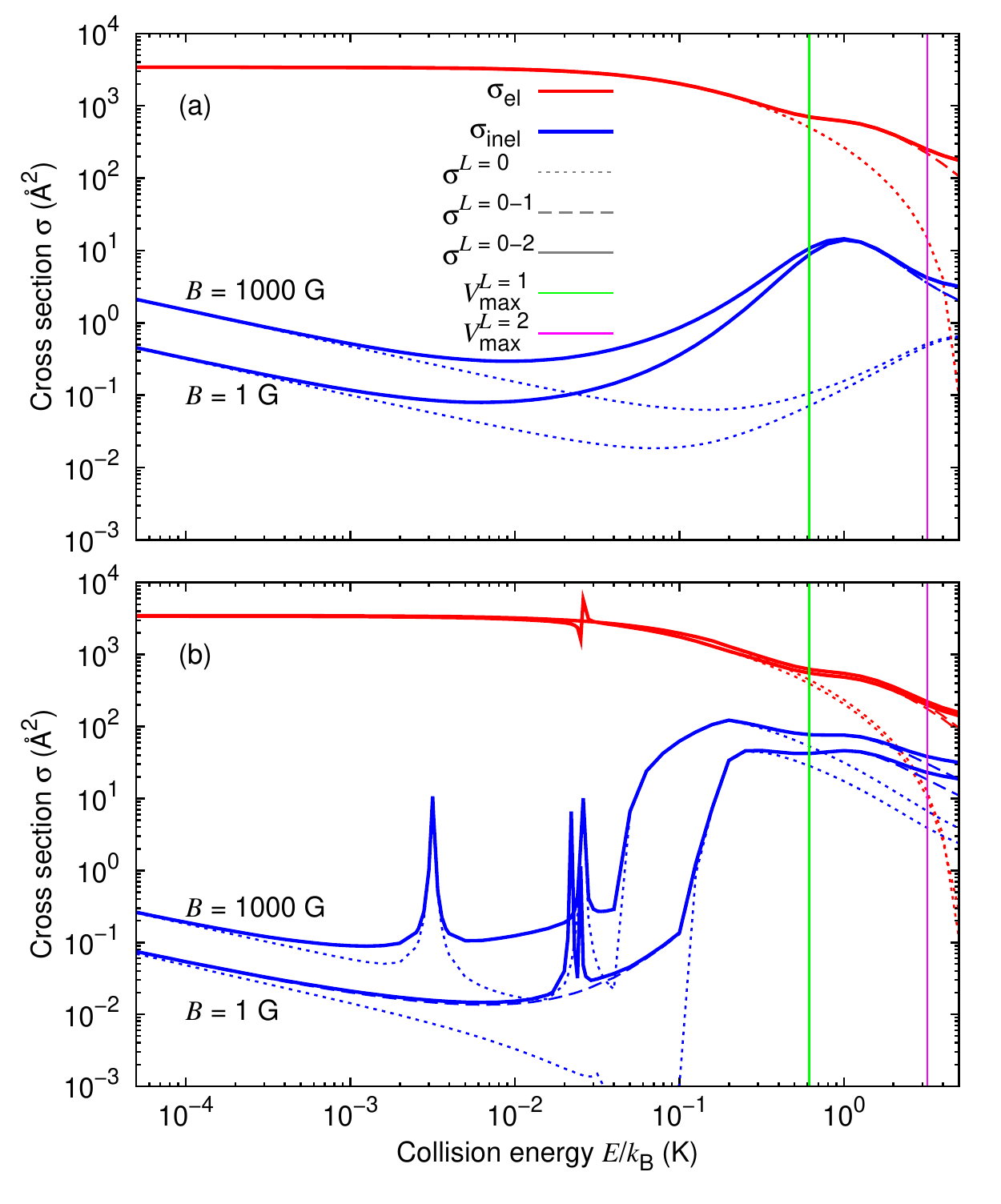}
\caption{(Color online) Elastic and total inelastic cross sections as a
function of collision energy, for $B=1$ and 1000~G.  Results are shown for
collisions of spin-stretched H (H$_d$) with: (a) spin-stretched F (F$_h$), and
(b) F$_c$ (the initial states highlighted in Fig.~\ref{fig:elevels}).  Solid
lines include s-, p- and d-wave contributions, dashed lines include s-, p-wave
contributions and dotted lines are the s-wave cross sections.  The vertical
lines show the heights of the p and d-wave centrifugal barriers.
 \label{fig:xsecE_B1+1000G}}
\end{figure}

\subsection{Elastic and inelastic collisions}
\label{sec3:collisions}

Figure \ref{fig:xsecE_B1+1000G}(a) shows the elastic and total inelastic cross
sections as a function of collision energy for collisions between hydrogen
atoms in state H$_d$ and fluorine atoms in state F$_h$. These collisions are
spin-stretched, meaning that both atoms have their maximum values of all
projection quantum numbers. Because of this, collisions that change a
projection quantum number on either atom must necessarily change $M_L$, and
this means that s-wave collisions (incoming $L=0$) must have outgoing $L\ge2$
\cite{AVAvdeenkov:01a, AVolpi:02a}. s-wave inelastic collisions are therefore
suppressed by an $L=2$ centrifugal barrier in the outgoing channel, which for
H+F collisions is 3.2~K high. For p-wave collisions, with incoming $L=1$, the
barrier is lower (614~mK), but in this case there is a barrier in both the
incoming and outgoing channels. Figure \ref{fig:xsecE_B1+1000G}(a) shows that
the elastic cross section is dominated by s-wave collisions at energies up to
about 400~mK. p-wave contributions to inelastic cross sections are significant
above about 1 mK and d-wave collisions contribute significantly above about
3~K. The low-energy inelastic cross sections are larger at 1000~G than at 1~G
because the kinetic energy release for inelastic collisions increases with
field and helps to overcome the centrifugal barrier in the outgoing channel.
Nevertheless, the ratio of elastic to inelastic cross sections remains above
100 for both fields at energies up to 5~K, except in the region of a p-wave
shape resonance in the incoming channel around 1~K.

\begin{figure}[t]
\includegraphics[width=85mm]{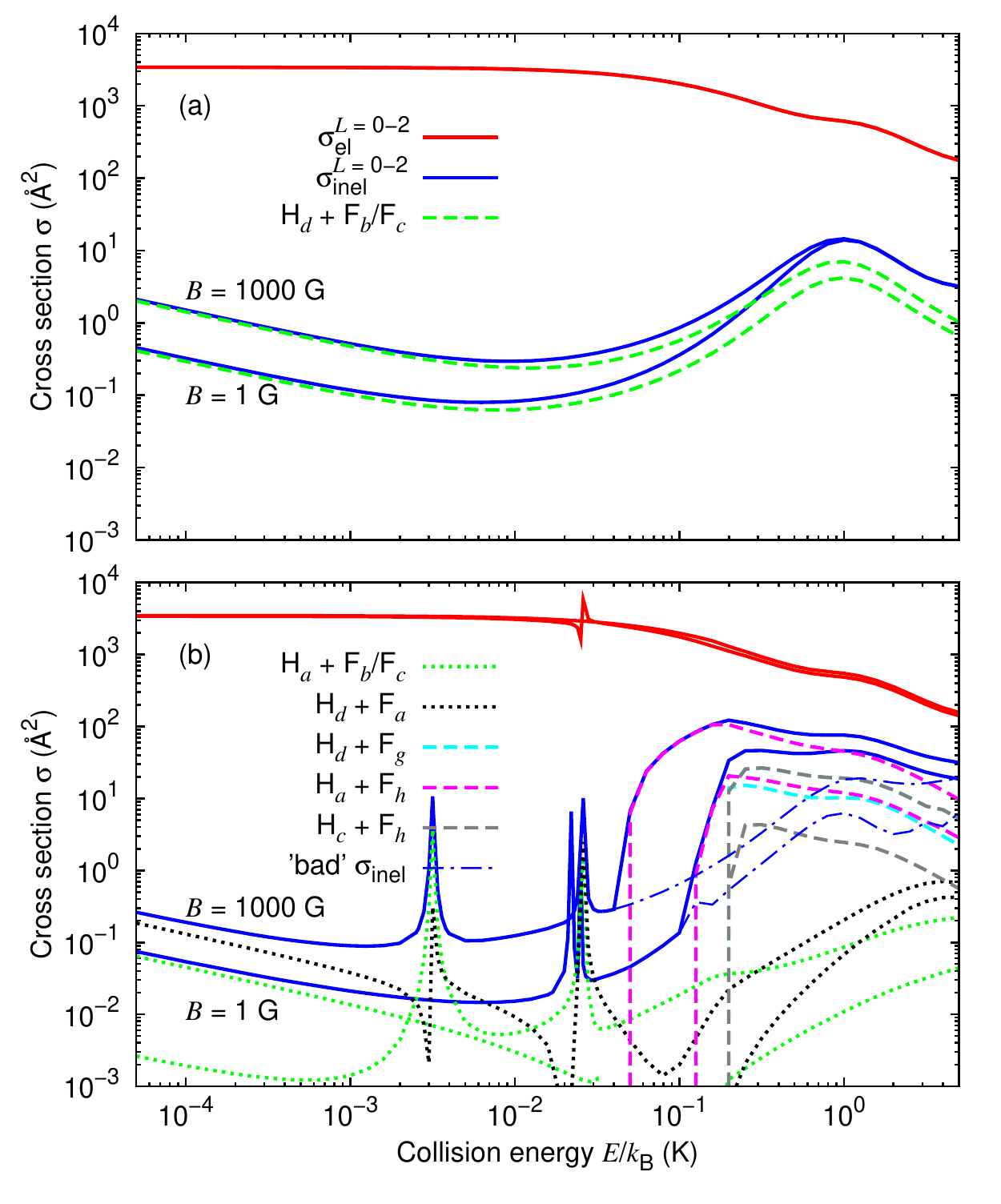}
\caption{(Color online) The most important state-to-state cross sections as a
function of collision energy, for $B=1$ and 1000~G.  Results are shown for
collisions of spin-stretched H (H$_d$) with: (a) spin-stretched F (F$_h$), and
(b) F$_c$. The line for H$_d$+F$_g$ is hidden underneath that for H$_c$+F$_h$
in the steep region near their thresholds. \label{fig:xsecE_B1+1000G_mech}}
\end{figure}

Figure \ref{fig:xsecE_B1+1000G_mech}(a) shows state-to-state cross sections for
the most important product channels in H$_d$ + F$_h$ collisions, which help to
understand the collision mechanisms. For spin-stretched collisions, the
dominant channels are those in which the F state changes and the H atom is a
spectator. This suggests that the dominant coupling is the anisotropy of the
triplet potentials $V^1_2(R)$ (Eq.\ \eqref{eq:VS02}), combined with the
spin-orbit coupling. This term can change $m_{f\mathrm{F}}$ by up to 2, and
formation of F$_b$ and F$_c$ is preferred over formation of F$_f$ and F$_g$
because the larger kinetic energy release for the former pair helps reduce
centrifugal suppression.

Figure \ref{fig:xsecE_B1+1000G}(b) shows elastic and total inelastic cross
sections for H$_d$ + F$_c$ collisions, with the F atom initially in a
magnetically trappable state that correlates with $f_\mathrm{F} = 1$; this
state is {\em not} spin-stretched and has $m_{f\mathrm{F}} = +1$. At 1000~G the
dominant channels at low collision energy are those in which the state of the H
atom does not change, as shown in Fig.\ \ref{fig:xsecE_B1+1000G_mech}(b). The
inelastic cross sections are actually {\em smaller} than for spin-stretched
collisions at low energy, because the kinetic energy releases are lower and
there is more centrifugal suppression.  At 1~G the centrifugal suppression is
so strong that a different mechanism takes over: the weak dipolar interaction
can drive long-range inelastic collisions that take place {\em outside} the
centrifugal barrier \cite{LMCJanssen:11a, LMCJanssen:11c}, producing H$_a$ and
F$_b$ or F$_c$; Fig.\ \ref{fig:xsecE_B1+1000G_mech}(b) shows that this is the
dominant low-energy mechanism at 1~G.

There are also narrow resonances for H$_d$ + F$_c$ collisions, between 20 and
30~mK at 1~G and around 3 and 30~mK at 1000~G, which produce peaks in the
inelastic cross sections. These are due to resonances in the singlet states and
their positions are very sensitive to changes in the singlet potentials.
However, they are narrow enough to have little effect on sympathetic cooling.

At higher energies, additional inelastic channels open. Collisions to form
H$_a$ + F$_h$ are possible at collision energies above 60~mK at 1~G and above
24~mK at 1000~G. There are also channels forming H$_c$ + F$_h$ and H$_d$ +
F$_g$ that open near 100~mK for both fields. All these collisions conserve
$m_{f\mathrm{F}}+m_{f\mathrm{H}}$, so are not centrifugally suppressed and
dominate the inelasticity above their threshold energies. They are driven by
the {\em difference} between the singlet and triplet potentials, and we refer
to them as ``spin-exchange" collisions by analogy with the corresponding
process in collisions between alkali-metal atoms. These collisions and their
consequences will be considered in more detail below. However, we will note
here that they do {\em not} lead to trap loss for F atoms, and will not
necessarily prevent sympathetic cooling; the dot-dashed lines in Fig.\
\ref{fig:xsecE_B1+1000G_mech}(b) show the inelastic cross sections obtained if
the spin-exchange collisions are excluded.

\begin{figure}[t]
 \includegraphics[width=85mm]{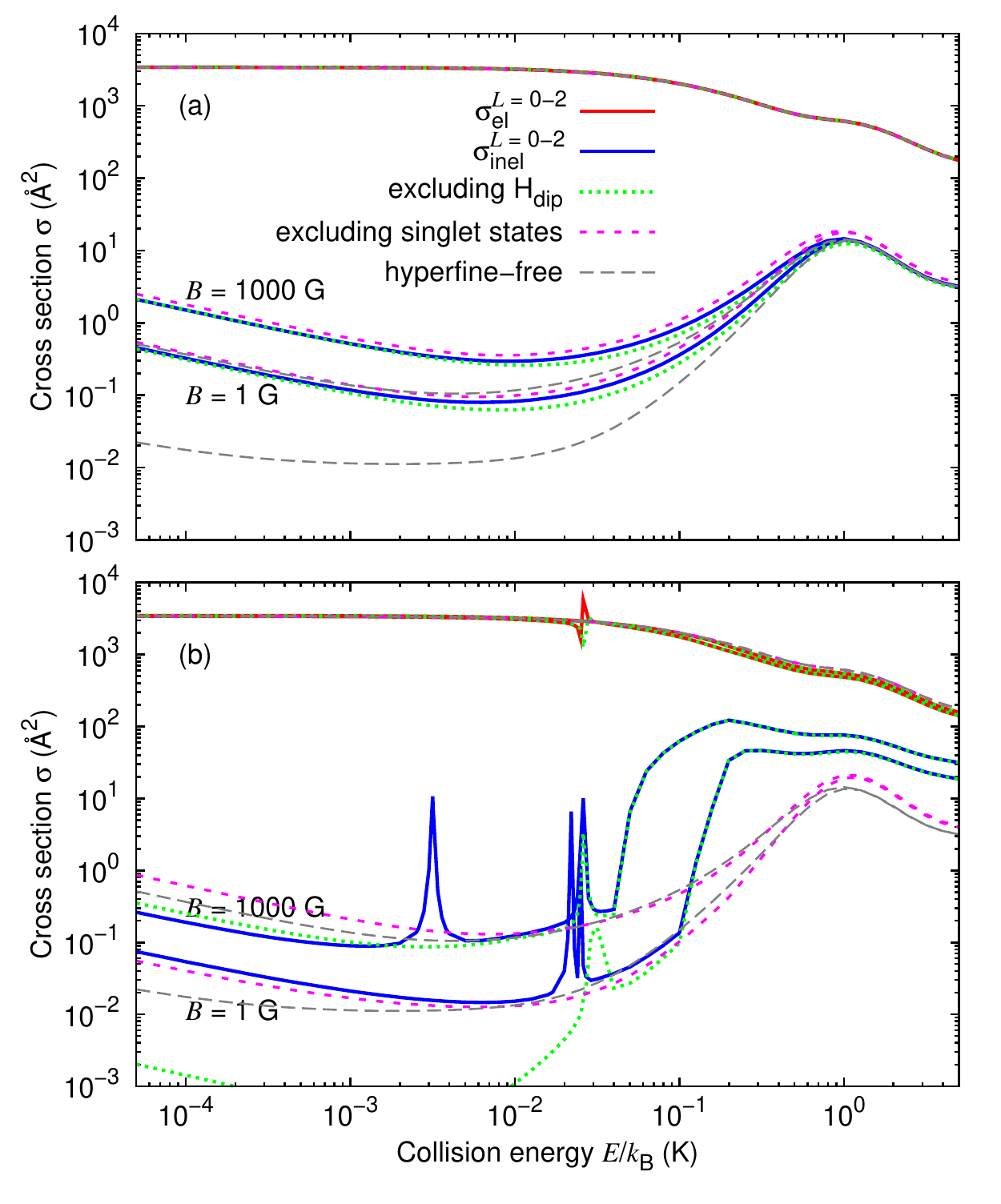}
 \caption{(Color online) Elastic and total inelastic cross sections as a
 function of collision energy, for $B=1$ and 1000~G, from both full
 coupled-channel calculations
 and a variety of approximations. Results are shown for collisions of
 spin-stretched H (H$_d$) with: (a) spin-stretched F (F$_h$), and (b) F$_c$.
 \label{fig:xsecE_B1+1000G_approx}}
\end{figure}

Figure \ref{fig:xsecE_B1+1000G_approx} shows the results for H$_d$ + F$_h$ and
H$_d$ + F$_c$ collisions with various approximations. First, the green dotted
lines show the results of excluding the dipolar interaction $\mathcal{H}_{\rm
dip}$ between the atoms; it may be seen that this is a good approximation for
spin-stretched collisions, and also for non-spin-stretched collisions at
1000~G, where the kinetic energy release is substantial and collisions that
change only $m_{f\mathrm{F}}$ dominate (driven by the potential anisotropy, and
subject to centrifugal suppression). However, it dramatically underestimates
the low-energy inelastic cross sections at 1~G, which are driven by the dipolar
interaction as discussed above.

The long dashed grey lines in Fig.\ \ref{fig:xsecE_B1+1000G_approx} show the
effect of neglecting hyperfine interactions (which produces the same state for
F$_h$ and F$_c$, with $m_{j\mathrm{F}} = +3/2$, so the hyperfine-free lines are
the same in both panels). This approximation significantly reduces the
inelastic cross sections: the difference arises from the additional kinetic
energy released in the presence of hyperfine interactions. This effect is
similar to the one that we have identified and discussed in Mg+NH collisions
\cite{MLGonzalez-Martinez:11}.

Lastly, the short dashed purple lines in Fig.\ \ref{fig:xsecE_B1+1000G_approx}
show the effect of setting the singlet potentials equal to the corresponding
triplet potentials. This approximation is not necessary for H+F collisions, but
for atom+molecule collisions such as H+NH and H+OH
\cite{MLGonzalez-Martinez:13a} it is difficult or impossible to converge
scattering calculations in which the deep low-spin surfaces (corresponding to
H$_2$O or NH$_2$) are included with their full depth. H+F allows this
approximation to be tested. It may be seen that including the full singlet
curves has only a fairly small effect (20 to 30\%) for spin-stretched
collisions; this is expected, because pairs of spin-stretched atoms or
molecules interact entirely on the highest-spin surface, and lower-spin
surfaces can contribute only after one of the spin projections has changed. For
the non-spin-stretched collisions shown in Fig.\
\ref{fig:xsecE_B1+1000G_approx}(b), the effect is small at 1~G, except near the
narrow singlet resonances, but considerably larger (up to a factor of 6) at
1000~G below 40 mK. In addition, neglecting the difference between the singlet
and triplet curves prevents the spin-exchange excitation collisions that become
important at higher collision energies.

\subsection{Sympathetic cooling} \label{sec2:SC}

Using a hard-sphere model, deCarvalho \emph{et al.}\ \cite{RdeCarvalho:99}
found that the temperatures difference between the warm and coolant species
after $k$ collisions is given by
\begin{equation}
\frac{T^k_2-T_1}{T^0_2-T_1} = \exp\left(-\frac{k}{\kappa}\right),
\end{equation}
where the number of collisions $\kappa$ required to reduce the temperature
difference by a factor of $e$ is
\begin{equation}
 \kappa = \frac{(m_1+m_2)^2}{2m_1m_2} = \frac{1}{2} \left(1+\frac{m_1}{m_2}\right)
 \left(1+\frac{m_2}{m_1}\right)
\end{equation}
and thus depends solely on the mass ratio. In our case, $\kappa \approx 10.5$.

\begin{figure}
 \includegraphics[width=85mm]{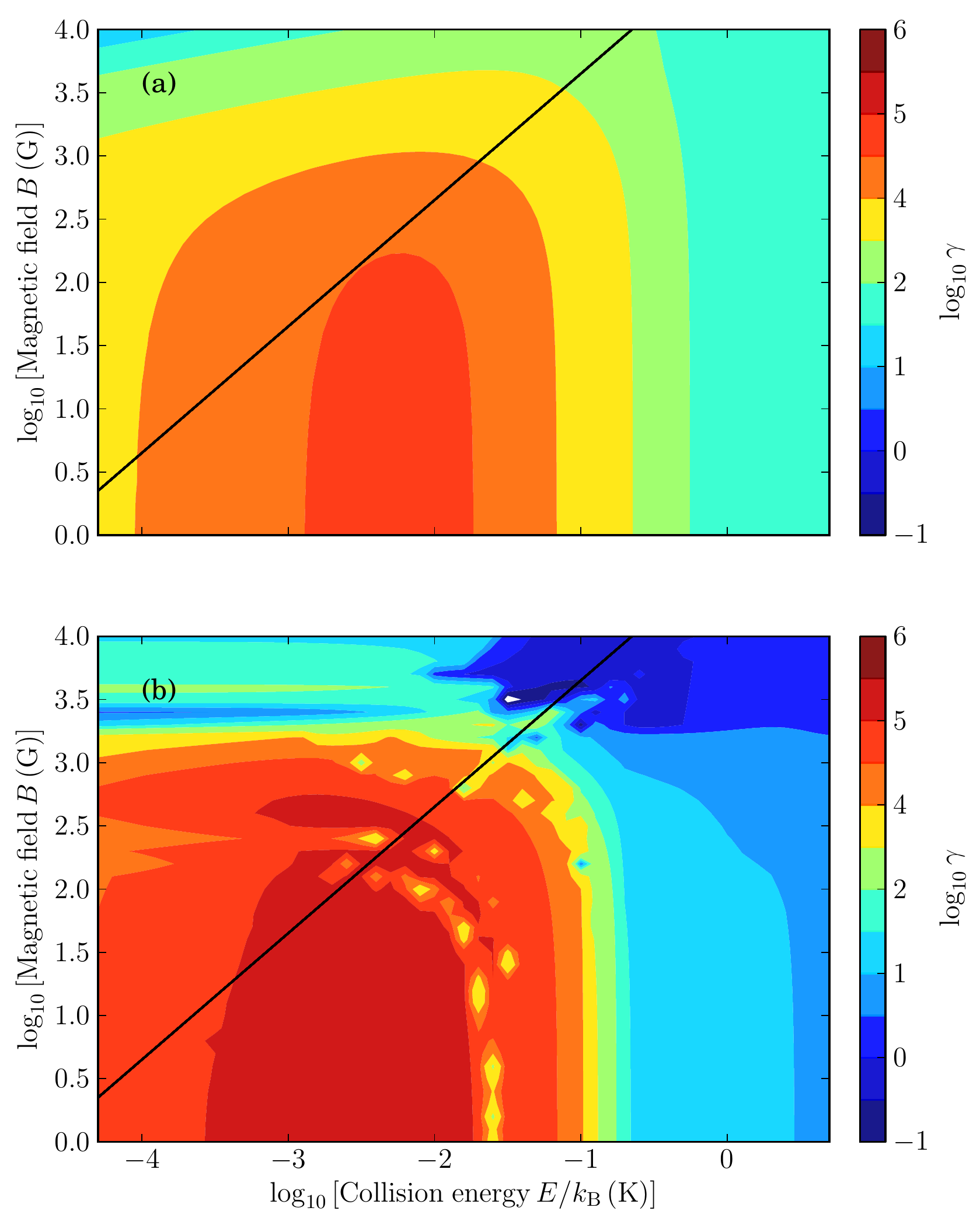}

\caption{(Color online) Contour plots of the ratio $\gamma$ of elastic to total
inelastic cross sections as a function of collision energy and magnetic field.
Results are shown for collisions of spin-stretched H (H$_d$) with: (a)
spin-stretched F (F$_h$), and (b) F$_c$ (the initial states highlighted in
Fig.~\ref{fig:elevels}). The apparently uneven behavior of the contours in (b)
arises because our grid cannot fully capture the sharp singlet resonances,
which on a finer grid would appear as very narrow bands rather than isolated
peaks.}
 \label{fig:gammaEB}
\end{figure}

Figure \ref{fig:gammaEB}(a) shows the ratio $\gamma$ of elastic to inelastic
cross sections, as a function of collision energy and magnetic field, for
$^{19}$F initially in state $\mathrm{F}_h$. The diagonal black line shows the
field at which the Zeeman energy is $6k_\mathrm{B}T$, so that 99.9\% of F atoms
at temperature $T$ sample fields below the line. It may be seen that $\gamma$
exceeds 1,000, and sympathetic cooling of $^{19}$F atoms is thus predicted to
be successful, for almost all fields and energies up to 5~K. The ratio reduces
to around 40 near 1~K, due to the p-wave shape resonance discussed above, but
this should not prevent overall cooling.

For F atoms initially in state F$_c$, the situation is rather more complicated.
Figure \ref{fig:gammaEB}(b) shows the ratio $\gamma$ for this state. At
energies and fields below about 50 mK, $\gamma$ for F$_c$ is actually more
favorable than for F$_h$. However, at higher energies the spin-exchange
collisions described above can occur, exciting the F atom to F$_h$ and
transferring the H atom to H$_a$. The latter is an untrapped state, so H atoms
in the H$_a$ state will be removed from the trap and will not be available for
the reverse process. However, the F atoms transferred to state F$_h$ will have
{\em lost} kinetic energy, so will remain trapped. If F atoms in the F$_c$
state start at a temperature of (say) 500~mK, it will require about 25 elastic
collisions to cool them to 50 mK (or somewhat more if the H atom temperature is
significant). However, the cross section for spin-exchange collisions is about
1/10 of that for elastic collisions, so almost all the atoms will be in state
F$_h$ by the time they have cooled to 50 mK. Sympathetic cooling will then
proceed mostly in state F$_h$ even if the atoms were initially in state F$_c$.

\subsection{Sensitivity to the interaction potential}

\begin{figure}
 \includegraphics[width=\linewidth]{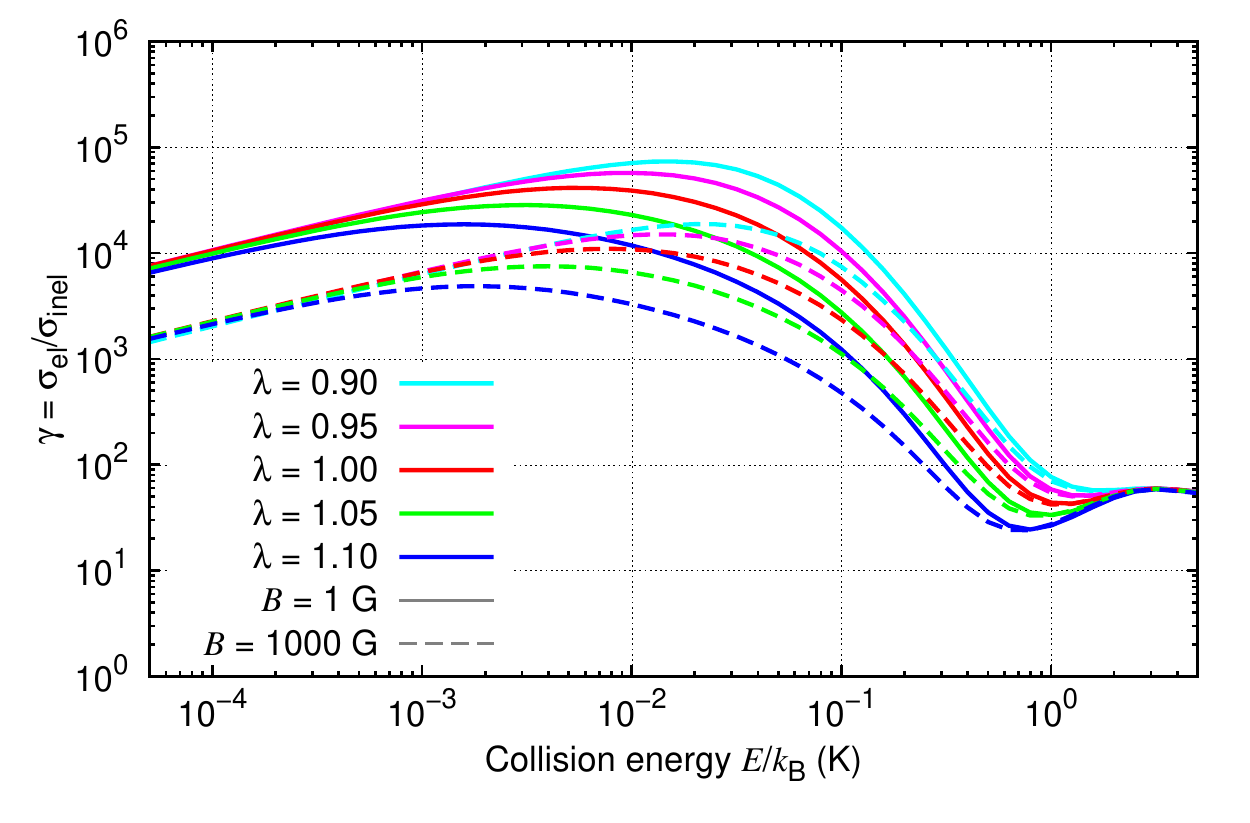}
 \caption{(Color online) Variation of the ratio $\gamma$ of elastic to total
 inelastic cross
 sections for spin-stretched collisions (H$_d$ + F$_h$) at 1~G (solid curves)
 and 1000~G (dashed curves), as a function of a scaling factor $\lambda$
 applied to the excited-state potential curves.
 \label{fig:gammaE_B1+1000G_lscan}}
\end{figure}

The ground-state potential for HF is very well known from spectroscopic
experiments \cite{JACoxon:12}, and as seen above has relatively little effect
on the inelastic cross sections, particularly for atoms in spin-stretched
states. However, there is some uncertainty in the potential curves for the
three excited states. To explore the effect of this uncertainty on the cross
sections, we have carried out calculations on sets of potentials obtained by
scaling the three excited curves by common factors of $\lambda=0.90$, 0.95,
1.05 and 1.10. This slightly changes the ${^1\Sigma^+}$ curve as well. The
resulting values of the ratio $\gamma$ for spin-stretched collisions at 1~G and
1000~G are shown in Fig.\ \ref{fig:gammaE_B1+1000G_lscan}. It may be seen that
$\gamma$ is almost independent of $\lambda$ below 1~mK, but that the
differences increase at higher energies. The main effect is that, as $\lambda$
is increased, the p-wave resonance near 1~K drops to slightly lower energies
and become higher and narrower. As seen in section \ref{sec2:SC}, the key
quantity is the energy at which the ratio $\gamma$ drops below 100: the worst
case is for $\lambda=1.10$ at 1000~G, where this is reduced from around 500 mK
to around 250 mK. We conclude that plausible variations in the potential may
slightly affect the temperature at which sympathetic cooling starts to work,
but do not alter the qualitative conclusions.

The ${^1\Sigma^+}$, ${^1\Pi}$, ${^3\Sigma^+}$ and ${^3\Pi}$ potential curves
used in the present work have scattering lengths of $-16.3$, $-107$, $-284$ and
12.1 \AA, respectively. These may be compared with the mean scattering length
$\bar{a}$ \cite{GFGribakin:93}, which for this system is 3.2~\AA. The isotropic
triplet potential $\hat{V}^1_0(R)$ supports one s-wave bound state just below
threshold. As $\lambda$ is reduced from 1, this state comes closer to
threshold; the s-wave elastic and inelastic cross sections both increase, but
by similar factors, so that $\gamma$ is almost unchanged at low energy. It
would require an implausibly small value $\lambda\approx0.8$ to bring this
state all the way up to threshold.

\section{Conclusions}
Ultracold hydrogen atoms are a very promising coolant for atoms and molecules
that are not amenable to laser Doppler cooling. In previous work
\cite{MLGonzalez-Martinez:13a}, we have shown that sympathetic cooling with
atomic hydrogen is likely to work for molecules such as NH and OH, from
starting temperatures of 100-1000 mK down to the microkelvin regime. In the
present paper we have shown that this is also true for atomic fluorine, from
starting temperatures of 500 to 1000 mK. Other halogen atoms are likely to
behave similarly, although more collisions are needed for thermalization
because of the high mass ratio.

The calculations on F+H collisions have also allowed us to test approximations
made for the molecular systems. In particular, for F+H it was possible to
include the deep singlet ground state fully in the calculations. Despite its
depth, the singlet state was found to have little effect on collisions
involving spin-stretched states. This gives us confidence that approximating
the deep low-spin surfaces, as was necessary for NH+H and OH+H
\cite{MLGonzalez-Martinez:13a}, is a good approximation.

\begin{acknowledgments} The authors are grateful to Dermot Green and Ruth
Le~Sueur for valuable discussions and to EPSRC for funding.
\end{acknowledgments}
\bibliographystyle{apsrev}
\bibliography{./refs_mlgm,../all}
\begin{widetext}
\appendix
\section{Matrix Elements}
\label{app:matrix_elements} This Appendix describes the matrix elements used in
the coupled-channel calculations in both the uncoupled and $ls$-coupled basis
sets described in Section \ref{sec2:basis+aHa'}.

The matrix elements for $\hat{L}^2$ are totally diagonal in both basis sets and
are given by $L(L+1)$.

Several terms in our Hamiltonian \eqref{eq:Heff} share the structure
$\hat{\mathcal{H}}_\mathrm{j_1j_2} = \kappa \hat{\jmath}_1 \cdot
\hat{\jmath}_2$, where $\kappa$ is a scalar while $\hat{\jmath}_1$ and
$\hat{\jmath}_2$ are vector operators.  Their matrix elements in the basis set
$|j_1 m_{j1}\rangle |j_2 m_{j2}\rangle$, are
 \begin{eqnarray}
  &&\hspace{-25mm}\langle j_2 m_{j2}| \langle j_1 m_{j1}|
   \hat{\mathcal{H}}_\mathrm{j_1j_2}
  |j_1 m'_{j1}\rangle |j_2 m'_{j2}\rangle =
   \delta_{m_{j1} m'_{j1}} \delta_{m_{j2} m'_{j2}} \kappa\, m_{j1} m_{j2}
   \nonumber\\ && +
   \delta_{m_{j1} m'_{j1} \pm 1} \delta_{m_{j2} m'_{j2} \mp 1} \frac{\kappa}{2}
   \left[j_1(j_1 + 1) - m_{j1} m'_{j1} \right]^{1/2}
   \left[j_2(j_2 + 1) - m_{j2} m'_{j2} \right]^{1/2}.
  \label{eq:aHj1j2a'}
 \end{eqnarray}
Such terms can mix functions with adjacent values of the projections of
$\hat{\jmath}_1$ and $\hat{\jmath}_2$, but preserve the sum $m_{12} = m_{j1} +
m_{j2}$. Here and throughout this appendix, the matrix elements are fully
diagonal with respect to quantum numbers that do not explicitly appear in their
definitions.

The matrix elements of $\hat{\mathcal{H}}_1$ depend on $|\alpha_1\rangle$ only
and are thus identical in the two basis sets.  Those for
$\hat{\mathcal{H}}_{\mathrm{F},1}$ are of the form \eqref{eq:aHj1j2a'}, while
those of $\hat{\mathcal{H}}_{\mathrm{Z},1}$ are totally diagonal and are given
by
 \begin{equation}
  \langle s_1 m_{s1}| \langle i_1 m_{i1}|
   \hat{\mathcal{H}}_\mathrm{Z,1}
  |i_1 m_{i1}\rangle |s_1 m_{s1}\rangle
  = \left(g_S\mu_\mathrm{B}m_{s1} - g_{i1}\mu_\mathrm{N}m_{i1}\right)B.
  \label{eq:aHZ1a'}
 \end{equation}

The matrix elements of $\hat{\mathcal{H}}_2$ depend on $|\alpha_2\rangle$ only.
Those for $\hat{\mathcal{H}}_\mathrm{so}$ are independent of the nuclear spin
quantum numbers.  They have the form \eqref{eq:aHj1j2a'} in the uncoupled basis
set, while in the $ls$-coupled basis set they are completely diagonal and are
given by
 \begin{equation}
   \langle (l_2 s_2)j_2 m_{j2}| \hat{\mathcal{H}}_\mathrm{so}
   |(l_2 s_2)j_2 m_{j2}\rangle
   = \frac{a_\mathrm{so}}{2} \left[j_2(j_2+1) - l_2(l_2+1) - s_2(s_2+1)\right].
  \label{eq:cHsoc'}
 \end{equation}
The matrix elements for the dipolar component of
$\hat{\mathcal{H}}_\mathrm{hf,2}$ are of the form \eqref{eq:aHj1j2a'} in both
basis sets (once $\hat{\jmath}_2$ is split into $\hat{l}_2 + \hat{s}_2$ for
evaluation in the uncoupled basis set).

The matrix elements for the quadrupolar term are readily calculated by first
rearranging
\begin{eqnarray}
 \langle\alpha_2| \hat{\mathcal{H}}_\mathrm{Q}| \alpha'_2\rangle
 = 2b_j \bigg(
 2\sum_{\alpha''_2}
  \langle\alpha_2| \hat\imath_2 \cdot \hat\jmath_2 |\alpha''_2\rangle
  \langle\alpha''_2| \hat\imath_2 \cdot \hat\jmath_2 |\alpha'_2\rangle
 + \langle\alpha_2| \hat\imath_2 \cdot \hat\jmath_2 |\alpha'_2\rangle \bigg)
 \label{eq:aHQa'}
\end{eqnarray}
and then applying Eq.~\eqref{eq:aHj1j2a'} to each term. For instance, in the
$ls$-coupled basis set
\begin{eqnarray}
 &&\hspace{-13mm}\langle i_2 m_{i2}| \langle (l_2 s_2)j_2 m_{j2}|
  \hat{\mathcal{H}}_\mathrm{Q}
 |(l_2 s_2)j_2 m'_{j2}\rangle |i_2 m'_{i2}\rangle = 2b_j \bigg\{
  \delta_{m_{i2} m'_{i2}} \delta_{m_{j2} m'_{j2}} 3 m_{i2} m_{j2}
  \nonumber\\ && +
  \delta_{m_{i2} m'_{i2} \pm 1} \delta_{m_{j2} m'_{j2} \mp 1} \frac{1}{2}
  \left(2m_{i2}m_{j2} + 2m'_{i2}m'_{j2} + 1\right)
  \left[i_2(i_2 + 1) - m_{i2} m'_{i2} \right]^{1/2}
  \left[j_2(j_2 + 1) - m_{j2} m'_{j2} \right]^{1/2}
  \nonumber\\ && + \frac{1}{2}
  \sum_{m''_{i2} m''_{j2}}
  \delta_{m_{i2} m''_{i2} \pm 1} \delta_{m_{j2} m''_{j2} \mp 1}
  \left[i_2(i_2 + 1) - m_{i2} m''_{i2} \right]^{1/2}
  \left[j_2(j_2 + 1) - m_{j2} m''_{j2} \right]^{1/2}
  \nonumber\\ && \times
  \delta_{m''_{i2} m'_{i2} \pm 1} \delta_{m''_{j2} m'_{j2} \mp 1}
  \left[i_2(i_2 + 1) - m''_{i2} m'_{i2} \right]^{1/2}
  \left[j_2(j_2 + 1) - m''_{j2} m'_{j2} \right]^{1/2} \bigg\},
 \label{eq:cHQc'}
\end{eqnarray}
which is non-zero for $\Delta m_{i2} = 0, \pm 1, \pm 2$ and $\Delta m_{j2} =
-\Delta m_{i2}$, thus preserving the sum $m_{i2} + m_{j2}$. The uncoupled
matrix elements are evaluated similarly by first splitting $\hat{\jmath}_2$
into $\hat{l}_2 + \hat{s}_2$.

The matrix elements for $\hat{\mathcal{H}}_\mathrm{Z,2}$ are totally diagonal
in the uncoupled basis set,
 \begin{equation}
   \langle s_2 m_{s2}| \langle l_2 m_{l2}| \langle i_2 m_{i2}|
    \hat{\mathcal{H}}_\mathrm{Z,2}
   |i_2 m_{i2}\rangle |l_2 m_{l2}\rangle |s_2 m_{s2}\rangle
   = \left(g'_L\mu_\mathrm{B}m_{l2} + g_S\mu_\mathrm{B}m_{s2} -
           g_{i2}\mu_\mathrm{N}m_{i2}\right)B,
  \label{eq:dHZ2d'}
 \end{equation}
while in the $ls$-coupled basis set they are given by
 \begin{eqnarray}
  &&\hspace{-13.5mm}
   \langle (l_2 s_2)j_2 m_{j2}| \langle i_2 m_{i2}|
    \hat{\mathcal{H}}_\mathrm{Z,2}
   |i_2 m_{i2}\rangle |(l_2 s_2)j'_2 m_{j2}\rangle
   \nonumber\\
   &=&
   \left[
    g_S \mu_\mathrm{B} (-1)^{2j_2+l_2+s_2-m_{j2}+1}
    \left[s_2(s_2+1)(2s_2+1)(2j_2+1)(2j'_2+1)\right]^{1/2}
    \left(\begin{array}{ccc} j_2    & 1 & j'_2   \\
                            -m_{j2} & 0 & m_{j2} \end{array}\right)
    \left\{\begin{array}{ccc} s_2 & j'_2 & l_2 \\
                              j_2 & s_2  & 1   \end{array}\right\}
   \nonumber\right.\\
   && +
    g'_L \mu_\mathrm{B} (-1)^{j_2+j'_2+l_2+s_2-m_{j2}+1}
    \left[l_2(l_2+1)(2l_2+1)(2j_2+1)(2j'_2+1)\right]^{1/2}
    \left(\begin{array}{ccc} j_2    & 1 & j'_2   \\
                            -m_{j2} & 0 & m_{j2} \end{array}\right)
    \left\{\begin{array}{ccc} l_2 & j'_2 & s_2 \\
                              j_2 & l_2  & 1   \end{array}\right\}
   \nonumber\\
   && -g_{i2}\mu_\mathrm{N}m_{i2}
   \bigg] B,
  \label{eq:cHZ2c'}
 \end{eqnarray}
where $\left(:::\right)$ and $\left\{:::\right\}$ represent Wigner 3-$j$ and
6-$j$ symbols.  $\hat{\mathcal{H}}_\mathrm{Z,2}$ can thus mix states with
different $j_2$ while preserving the projection $m_{j2}$: for fluorine, it
couples the $j_2 = 3/2$ and $1/2$ states only for $m_{j2} = -1/2$ or $1/2$.

The matrix elements of the electronic interaction potential in the uncoupled
basis set are
 \begin{eqnarray}
  &&\hspace{-7.5mm}
   \langle L M_L| \langle s_2 m_{s2}| \langle l_2 m_{l2}| \langle s_1 m_{s1}|
    \hat{V}
   |s_1 m'_{s1}\rangle |l_2 m'_{l2}\rangle |s_2 m'_{s2}\rangle |L' M'_L\rangle
   \nonumber\\
   &=& (-1)^{2(s_1-s_2+M_S)-m_{l2}+M'_L} (2l_2+1)
   \left[(2L+1)(2L'+1)\right]^{1/2}
   \sum_S (2S+1)
    \left(\begin{array}{ccc} s_1    & s_2     & S    \\
                            m_{s1}  & m_{s2}  & -M_S \end{array}\right)
    \left(\begin{array}{ccc} s_1     & s_2     & S   \\
                            m'_{s1} & m'_{s2} & -M_S \end{array}\right)
   \nonumber\\
   && \times
   \sum_{k} \hat{V}^S_k(R)
    \left(\begin{array}{ccc} l_2 & k & l_2 \\
                             0   & 0 & 0   \end{array}\right)
    \left(\begin{array}{ccc} l_2    & k   & l_2     \\
                            -m_{l2} & m_k & m'_{l2} \end{array}\right)
    \left(\begin{array}{ccc} L & k & L' \\
                             0 & 0 & 0  \end{array}\right)
    \left(\begin{array}{ccc} L   & k    & L'   \\
                            -M_L & -m_k & M'_L \end{array}\right),
  \label{eq:dVd'}
 \end{eqnarray}
with $M_S \equiv m_{s1}+m_{s2} =  m'_{s1}+m'_{s2}$ and $m_k \equiv M'_L-M_L$.
The matrix elements for the isotropic terms $V^S_0$ are off-diagonal only in
the $m_s$ quantum numbers while preserving $M_S$.  If $l_2 = 0$, these are the
only couplings induced by $\hat V$ and have been studied in detail in
collisions of alkali atoms.  For $l_2 \ne 0$, the anisotropic terms $V^S_k$ ($k
\ne 0$) can additionally mix different partial waves and change $m_{l2}$
independently of $m_{s2}$, thus changing their sum $m_{j2}$ (such couplings
exist even if $s_1 = 0$).

The mixing of the electronic degrees of freedom makes it difficult to find an
expression for the matrix elements of the electronic interaction in the
$ls$-coupled basis set that has a simple physical interpretation.  It is more
convenient to evaluate these by transforming the matrix elements in the
uncoupled representation \eqref{eq:dVd'} using the standard vector-coupling
formula,
 \begin{equation}
  |(l_2 s_2)j_2 m_{j2}\rangle = \sum_{m_{l2} m_{s2}}
    \langle l_2 m_{l2} s_2 m_{s2}|(l_2 s_2)j_2 m_{j2}\rangle
    |l_2 m_{l2}\rangle |s_2 m_{s2}\rangle.
 \end{equation}

If the extremely small nuclear contributions to the magnetic moments in
$\hat{\mathcal{H}}_\mathrm{dip}$ are neglected, the matrix elements of
$\hat{\mathcal{H}}_{12}$ become diagonal in the nuclear projection quantum
numbers. The matrix elements of the dipolar interaction in the uncoupled basis
set are given by
 \begin{eqnarray}
  &&\hspace{-11.2mm}
   \langle L M_L| \langle s_2 m_{s2}| \langle l_2 m_{l2}| \langle s_1 m_{s1}|
    \hat{\mathcal{H}}_\mathrm{dip}
   |s_1 m'_{s1}\rangle |l_2 m'_{l2}\rangle |s_2 m'_{s2}\rangle |L' M'_L\rangle
   \nonumber\\
   &=& -\sqrt{30}
   \delta_{m_{l2} m'_{l2}} \lambda_{s1s2}(R) (-1)^{s_1-m_{s1}+s_2-m_{s2}-M_L}
   \left[s_1(s_1+1)(2s_1+1) s_2(s_2+1)(2s_2+1) (2L+1)(2L'+1)\right]^{1/2}
   \nonumber\\
   && \times
   \left(\begin{array}{ccc} L & 2 & L' \\
                            0 & 0 & 0  \end{array}\right)
   \sum_{q_1,q_2}
    \left(\begin{array}{ccc} 1   & 1   & 2          \\
                             q_1 & q_2 & -q         \end{array}\right)
    \left(\begin{array}{ccc} s_1    & 1   & s_1     \\
                            -m_{s1} & q_1 & m'_{s1} \end{array}\right)
    \left(\begin{array}{ccc} s_2    & 1   & s_2     \\
                            -m_{s2} & q_2 & m'_{s2} \end{array}\right)
    \left(\begin{array}{ccc} L   & 2   & L'         \\
                            -M_L & -q & M'_L        \end{array}\right)
   \nonumber\\
   && -\sqrt{30}
   \delta_{m_{s2} m'_{s2}} \lambda_{s1l2}(R) (-1)^{s_1-m_{s1}+l_2-m_{l2}-M_L}
   \left[s_1(s_1+1)(2s_1+1) l_2(l_2+1)(2l_2+1) (2L+1)(2L'+1)\right]^{1/2}
   \nonumber\\
   && \times
   \left(\begin{array}{ccc} L & 2 & L' \\
                            0 & 0 & 0  \end{array}\right)
   \sum_{q_1,q_2}
    \left(\begin{array}{ccc} 1   & 1   & 2  \\
                             q_1 & q_2 & -q \end{array}\right)
    \left(\begin{array}{ccc} s_1    & 1    & s_1     \\
                            -m_{s1} & q_1 & m'_{s1} \end{array}\right)
    \left(\begin{array}{ccc} l_2    & 1   & l_2     \\
                            -m_{l2} & q_2 & m'_{l2} \end{array}\right)
    \left(\begin{array}{ccc} L   & 2  & L'   \\
                            -M_L & -q & M'_L \end{array}\right),
  \label{eq:dHdipd'}
 \end{eqnarray}
 with the $R$-dependent spin-spin coupling $\lambda_{s1s2}(R) =
 E_\mathrm{h}\alpha^2a^3_0/R^3$ ($\alpha\approx1/137$ is the fine-structure
 constant), $\lambda_{s1l2} = \left(g'_L/g_S\right) \lambda_{s1s2}$ and
 $q \equiv q_1 + q_2$.  In the $ls$-coupled basis set,
 \begin{eqnarray}
  &&\hspace{-7mm}
   \langle L M_L| \langle (l_2 s_2)j_2 m_{j2}| \langle s_1 m_{s1}|
    \hat{\mathcal{H}}_\mathrm{dip}
   |s_1 m'_{s1}\rangle |(l_2 s_2)j'_2 m'_{j2}\rangle |L' M'_L\rangle
   \nonumber\\
   &=& \sqrt{30}
   (-1)^{s_1-m_{s1}+j_2+l_2+s_2-m_{j2}-M_L}
   \left[s_1(s_1+1)(2s_1+1) (2j_2+1)(2j'_2+1) (2L+1)(2L'+1)\right]^{1/2}
   \left(\begin{array}{ccc} L & 2 & L' \\
                            0 & 0 & 0  \end{array}\right)
   \nonumber\\
   && \times
   \left[
    \lambda_{s1s2}(R) (-1)^{j_2} \left[s_2(s_2+1)(2s_2+1)\right]^{1/2}
    \left\{\begin{array}{ccc} s_2 & j'_2 & l_2 \\
                              j_2 & s_2  & 1   \end{array}\right\}
    +\lambda_{s1l2}(R) (-1)^{j'_2} \left[l_2(l_2+1)(2l_2+1)\right]^{1/2}
    \left\{\begin{array}{ccc} l_2 & j'_2 & s_2 \\
                              j_2 & l_2  & 1   \end{array}\right\}
   \right]
   \nonumber\\
   && \times
   \sum_{q_1,q_2}
    \left(\begin{array}{ccc} 1   & 1   & 2          \\
                             q_1 & q_2 & -q         \end{array}\right)
    \left(\begin{array}{ccc} s_1    & 1   & s_1     \\
                            -m_{s1} & q_1 & m'_{s1} \end{array}\right)
    \left(\begin{array}{ccc} j_2    & 1   & j_2     \\
                            -m_{j2} & q_2 & m'_{j2} \end{array}\right)
    \left(\begin{array}{ccc} L   & 2   & L'         \\
                            -M_L & -q & M'_L        \end{array}\right).
  \label{eq:cHdipc'}
 \end{eqnarray}
These expressions show that $\hat{\mathcal{H}}_\mathrm{dip}$ adds to the
couplings produced by $\hat V$ in that: (1) it relaxes the constraint of $M_S$
being conserved, as long as $\Delta M_S = -\Delta M_L$; and (2) it allows for
changes in $m_{l2}$ that preserve $M_L$. The 3-$j$ symbols in
Eqs.~\eqref{eq:dVd'}--\eqref{eq:cHdipc'} restrict $\Delta L$ to be even, and
thus channels with only even (or only odd) $L$ values appear in the basis set
for a given total parity $P$.
\end{widetext}

\end{document}